\documentclass[journal=jctc,manuscript=article,layout=twocolumn]{achemso}
\usepackage{achemso}
\setkeys{acs}{maxauthors = 0}

\SectionNumbersOn
\usepackage{amsmath}
\usepackage{graphicx}% Include figure files
\usepackage{dcolumn}% Align table columns on decimal point
\usepackage{bm}% bold math
\usepackage{hyperref}% add hypertext capabilities
\usepackage[capitalise]{cleveref}
\usepackage{makecell}
\usepackage{xcolor}   
\usepackage{wasysym}
\usepackage{tikzsymbols}
\usepackage{multirow}
\usepackage[outdir=./]{epstopdf}

\title{Accelerated Organic Crystal Structure Prediction with Genetic Algorithms and Machine Learning}% Force line breaks with \\

\author{Amit Kadan\thefootnote{}}%
\email{amit@goodchemistry.com}
\affiliation{Good Chemistry Company, 1285 W Pender St, Vancouver, BC, Canada, V6E 4B1}%
\altaffiliation{Contributed equally to this work}

\author{Kevin Ryczko\thefootnote{}}
 \email{kevin@goodchemistry.com}
\affiliation{Good Chemistry Company, 1285 W Pender St, Vancouver, BC, Canada, V6E 4B1}%
\altaffiliation{Contributed equally to this work}

\author{Andrew Wildman}
\affiliation{Good Chemistry Company, 1285 W Pender St, Vancouver, BC, Canada, V6E 4B1}%

\author{Rodrigo Wang}
\affiliation{Good Chemistry Company, 1285 W Pender St, Vancouver, BC, Canada, V6E 4B1}%

\author{Adrian Roitberg}
\email{roitberg@ufl.edu}
\affiliation{
University of Florida, Department of Chemistry, PO Box 117200, Gainesville, FL, USA 32611-7200 
}%

\author{Takeshi Yamazaki}
\email{takeshi@goodchemistry.com}
\affiliation{Good Chemistry Company, 1285 W Pender St, Vancouver, BC, Canada, V6E 4B1}%

\date{\today}% It is always \today, today,
\begin{document}

\begin{tocentry}
    \includegraphics[width=\linewidth]{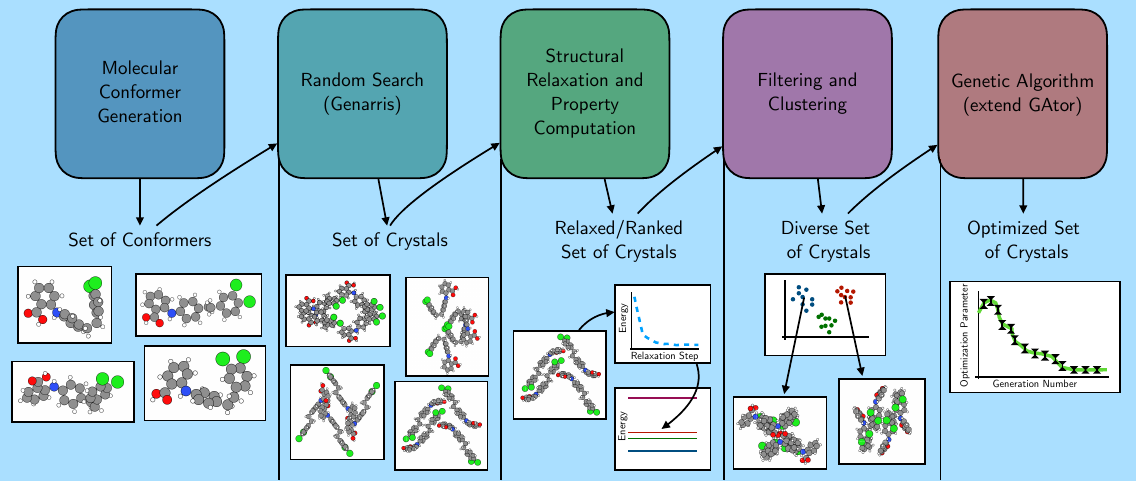}
\end{tocentry}

\begin{abstract}
We present a high-throughput, end-to-end pipeline for organic crystal structure prediction (CSP) -- the problem of identifying the stable crystal structures that will form from a given molecule based only on its molecular composition.
Our tool uses Neural Network Potentials (NNPs) to allow for efficient screening and structural relaxations of generated crystal candidates. 
Our pipeline consists of two distinct stages -- random search, whereby crystal candidates are randomly generated and screened, and optimization, where a genetic algorithm (GA) optimizes this screened population.
We assess the performance of each stage of our pipeline on 21 molecules taken from the Cambridge Crystallographic Data Centre's CSP blind tests. 
We show that random search alone yields matches for $\approx 50\%$ of targets. We then validate the potential of our full pipeline, making use of the GA to optimize the Root Mean-Squared Deviation (RMSD) between crystal candidates and the experimentally derived structure. With this approach, we are able to find matches for $\approx80\%$ of candidates with 10-100 times smaller initial population sizes than when using random search.
Lastly, we run our full pipeline with an ANI model that is trained on a small dataset of molecules extracted from crystal structures in the Cambridge Structural Database, generating $\approx 60\%$ of targets. By leveraging ML models trained to predict energies at the DFT level, our pipeline has the potential to approach the accuracy of \emph{ab initio} methods and the efficiency of empirical force-fields.
\end{abstract}

\maketitle

\section{Introduction}
When a compound crystallizes, the constituent molecules can arrange themselves in many different configurations. These multiple stable forms are called polymorphs, each of which may exhibit different chemical and physical properties \cite{bernstein2020polymorphism}. The problem of organic (or molecular) crystal structure prediction (CSP)  has important applications in a number of areas including drug discovery, where the crystallized form of a molecule may affect the efficacy, safety, and formulation of a drug~\cite{bauer2001ritonavir,day2007strategy,reilly2014role}, and in the development of organic semiconductors in flexible electronic devices, where polymorphism has consequences for opto-electronic performance~\cite{musil2018machine,reese2007organic,hasegawa2009organic,elder2015solid,cudazzo2015exciton}. In a broader context, the ability to do accurate CSP \textit{in-silico}, as a complimentary approach to experimental high-throughput screening in the laboratory, has shown proof of greatly reducing the time and money required to generate a candidate crystal with desired properties~\cite{chen2019investigation,omar2021high}, and has the potential to allow practitioners to explore a larger part of chemical space, facilitating the discovery of novel chemical entities~\cite{mcdonagh2019machine}.

The benefits of being able to do accurate CSP can be illustrated by the example of Ritonavir, an antiretroviral drug used to treat HIV, which has two known polymorphs~\cite{bauer2001ritonavir}. During its development, a single polymorph called form I was identified. After the drug went to market, a second, lower free energy polymorph -- form II, was discovered. This lower energy polymorph was significantly less likely to be absorbed into the bloodstream, compromising the efficacy of the drug. Even a trace amount of the second, more stable polymorph resulted in the conversion of form I to form II, leading to the recall of Ritonavir. It is estimated that the pharmaceutical company responsible for Ritnoavir lost around \$250 million USD over the incident \cite{buvcar2015disappearing}.

Despite its far-reaching effects, CSP remains an extremely difficult task to solve. The issues in CSP are twofold since one must pair an effective sampling routine with an efficient and accurate scoring method. Sampling is made difficult by the vast chemical space that must be searched through. Crystals exhibit high levels of symmetry, with 230 distinct space groups existing in three dimensions. Furthermore, the unit cell may come from one of 14 bravais lattice systems, each of which has different associated compatible space groups~\cite{aroyo2013international}. Scoring different crystal candidates is made difficult by the small energy differences between low-energy polymorphs, which rarely exceed 10 kJ/mol, requiring one to score structures with quantum mechanical potentials to accurately characterize differences in stability~\cite{nyman2015static}. In practice, DFT, in particular with the PBE0 exchange-correlation functional \cite{adamo1999toward,ernzerhof1999assessment} and the inclusion of dispersion effects, has shown to have good ranking ability for crystal systems~\cite{reilly2013understanding,marom2013many, kronik2014understanding, moellmann2014dft, reilly2016sixthblindtest,hoja2018first, price2023xdm}. However, DFT is expensive to evaluate, potentially requiring tens of thousands of hours to relax a single crystal system on a standard CPU~\cite{wengert2021data}. There is an intimate interplay between these two requirements. An accurate scoring function can guide the sampling routine to promising areas of the search space, while an effective sampling routine aids in decreasing the number of calls to the scoring function, reducing the cost associated with the discovery of novel polymorphs.

Due to the various applications of organic crystals and the challenge of predicting stable polymorphs given a particular molecule, the Cambridge Crystallographic Data Centre has periodically organized ``blind-test'' challenges \cite{lommerse2000test, motherwell2002crystal, day2005third, day2009significant, bardwell2011towards, reilly2016report}, where teams aim to predict the stable polymorph of a target molecule (derived through experiment) given only the chemical diagram of the molecule. While the first blind test focused on small rigid molecules, the most recent blind test contained more difficult targets which included bulky flexible molecules, multi-component crystals, and a former drug candidate which has 5 known polymorphs. These challenges have profoundly driven the field of organic CSP, with the 6th blind test receiving 25 submissions from almost 100 researchers \cite{reilly2016report}. In the blind tests, various methods were used to generate and rank crystal structures. Sampling techniques included random search, quasi-random search, grid search, genetic algorithms, Monte-Carlo simulated annealing, and Monte-Carlo parallel tempering. Ranking techniques included empirical potentials, atomic multipoles, and  density functional theory (DFT). 

In this work, we propose a novel approach to overcoming the two major hurdles of CSP by pairing an alternative ranking technique utilizing neural network potentials (NNPs) with a genetic algorithm (GA) to produce diverse, high-quality samples from structural space.
The use of machine learning allows one to approach the speed of an empirical potential while approaching the accuracy of the underlying lower-level theory on which the model is trained. To predict organic crystals one must potentially generate and rank millions of candidate crystals. This becomes extremely computationally demanding when using \textit{ab initio} methods. In the 6th blind test, hundreds of thousands of CPU hours were used to rank structures using DFT~\cite{reilly2016sixthblindtest}. By using machine learning models trained to predict energies at the DFT level of theory, one can achieve a similar accuracy with a significantly lowered computational cost \cite{smith2018less, smith2019outsmarting, devereux2020extending}. GAs have been successfully applied to problems in CSP~\cite{curtis2018gator, curtis2018evolutionary, bier2021crystal, falls2020xtalopt, tom2023inverse, lyakhov2013new, zhu2012constrained, abardeh2022crystal}.
GAs make improvements to a population of candidate solutions by using biologically inspired operators to modify individual candidates. GAs proceed in ``generations", whereby a selection rule balancing exploration with exploitation is used to select candidates for modification. Our GA uses a variety of operators and selection rules to balance the trade-off between producing diverse, and high-quality samples.

There have been several past reports that use machine learning to facilitate organic crystal structure prediction. In one report Wengert \textit{et al.}~\cite{wengert2021data} trained a $\Delta$-ML method to learn the gap between dispersion-corrected density functional tight binding (DFTB) \cite{brandenburg2014accurate, yang2007extension} and semi-local DFT (using the PBE functional \cite{perdew1996generalized}) with many body dispersion (DFT-MBD) \cite{tkatchenko2012accurate}. Although they show that this model accurately ranks several organic crystals, they only exhibit the generative power of their framework on a single candidate from the latest blind test. Moreover, since their method requires a DFTB calculation, the computational scaling is worse than a pure machine learning approach. In another report, Kilgour \textit{et al.}~\cite{kilgour2023geometric} trained a geometric deep learning model to rank organic crystals. However, they do not integrate their model into a corresponding sampling framework and do not support the evaluation of multi-component crystals. The ability to evaluate multi-component crystals is becoming increasingly important. This is reflected in the 6th blind test, where 2/5 targets corresponded to multi-component crystals.

\begin{figure*}[ht]
    \centering
    \includegraphics[width=\linewidth]{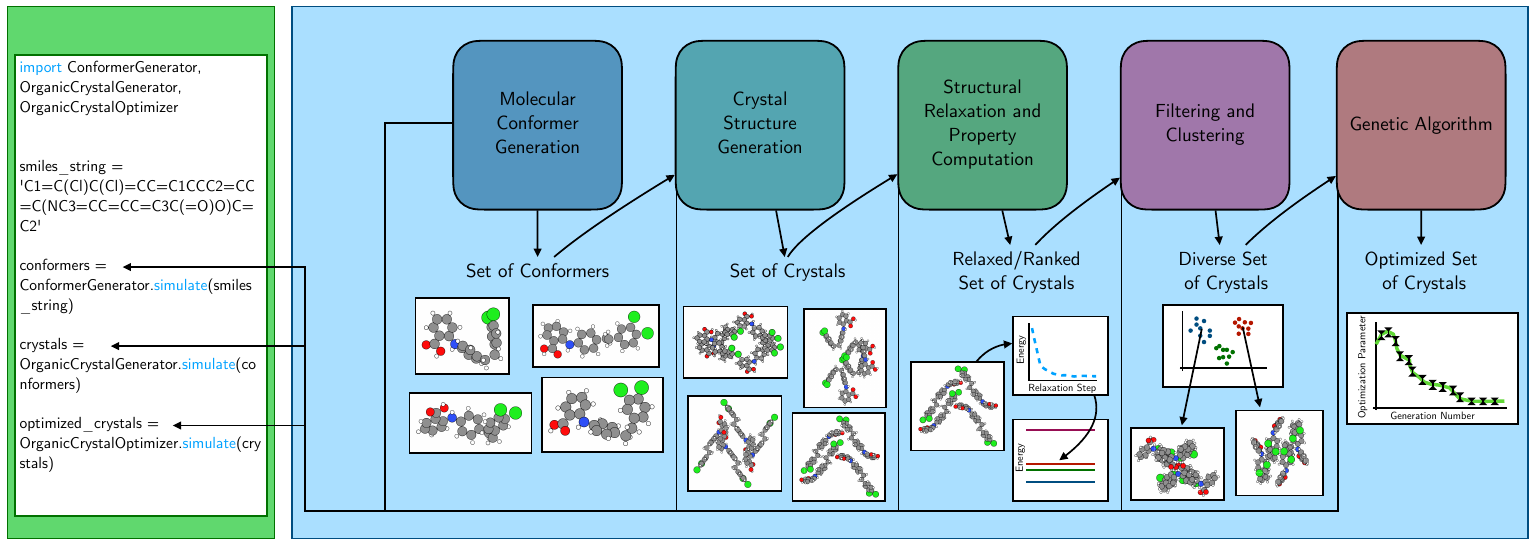}
    \caption{Visual representation of the organic crystal generation pipeline presented in this work. In a few lines of Python code, one can calculate a set of molecular conformers and a set of (optimized) organic crystals. }
    \label{fig:csp_overview}
\end{figure*}

Our pipeline, which we refer to as GAmuza (named after a mountain peak located in the North Cascades of southwestern British Columbia), combines multiple methods from past reports -- generating random organic crystals using Genarris \cite{tom2020genarris}, and optimizing these crystals using a novel genetic algorithm with breeding operations taken from GAtor \cite{curtis2018gator}. We use the ANI NNPs~\cite{smith2018less, smith2019outsmarting, devereux2020extending} throughout our pipeline to evaluate the lattice energy of generated structures. Within GAmuza, we introduce specialized organic crystal breeding operations that preserve the symmetry groups of the population. 

The organization of this report proceeds in the following manner; in \cref{methods}, we discuss the two stages of our pipeline -- random search, which includes conformer generation, random structure generation, local relaxation, filtering and clustering, and the GA which includes selection, breeding operations, and evolutionary niching. 

In \cref{results}, we assess the performance of the ANI models to rank crystal structures on molecules from past blind tests after generating large pools of structures via random search. 

Afterwards, we validate the GA by using it to optimize the RMSD between generated crystal candidates and the experimentally derived structures. We compare using the GA with all of the original operators included in GAtor versus using only symmetry-preserving operations. Lastly, we assess the performance of our entire pipeline when equipped with an ANI potential for both ranking during random search and selection and relaxation within the GA. We then compare the performance of the pipeline when used with the baseline ANI2x~\cite{devereux2020extending} model, and a variant which is transfer learned on a dataset of molecules extracted from organic crystals in the Cambridge Structural Database (CSD)~\cite{allen2002cambridge}. Lastly, in \cref{conclusion} we conclude and outline our future work based on our present results.

\section{Methods}
\label{methods}

In this section, we describe our entire pipeline which can be seen graphically in \cref{fig:csp_overview}. 

\subsection{Random Search}\label{subsec:random_gen}

\subsubsection{Conformer Generation}
 Our pipeline begins with a molecule described by a SMILES \cite{weininger1988smiles} string. 
 This SMILES string is used to generate 3D molecular conformers using a call to RDKit's \texttt{emebedMultipleConfs} \cite{landrum2013rdkit}. This generates multiple conformers for a given molecule, retaining only those where the root mean-squared deviation (RMSD) of heavy atoms after superimposing with other generated conformers is greater than some threshold RMSD value. We attempt to generate 1000 conformers per molecule, using the default RMSD threshold value (1 \AA) for retaining structurally dissimilar conformers.
 We found diminishing returns in terms of the number of structures retained when generating more than $1000$ initial conformers with RDKit and hence chose this number. The resulting conformers are then optimized using the MMFF force field within RDKit \cite{tosco2014bringing}. In the present study, the MMFF force field was chosen due to its ease of use within RDKit, and its ability to accurately describe organic compounds relative to other force fields~\cite{lewis2022comparing}. For multi-component crystals described by a SMILES string containing multiple molecules, we follow the same procedure when generating each molecule's individual conformers. To generate a joint conformer, we randomly place each of the molecules' conformers next to one other, ensure that no overlap exists between the two structures, and further optimize their joint structure using the MMFF force field. Conformers are then pruned such that the RMSD of heavy atoms between two configurations is never $<$ 1 \AA. In our future work, we plan to replace the force field optimizations with machine learning models trained on accurate quantum chemistry data (i.e. ANI) to optimize the coordinates of the molecular conformers as was done in Ref. \cite{liu2022auto3d}. This will enable accurate and rapid molecular conformer generation within GAmuza. If one chooses the ANI models to perform this task, one should be aware of their atom type limitation. To be able to handle organometallics, for example, one must generate consistent data in this area of the chemical space and retrain the model.

 \subsubsection{Initial Crystal Structure Generation}
 Given the set of conformers and a specified number of molecules per cell, we use Genarris \cite{tom2020genarris} to generate crystal structures. To describe the Genarris algorithm briefly, first, the volume of the cell is estimated using the van der Waals (vdW) radii of atoms. The number of compatible space groups is then identified given the requested number of molecules per cell. 

 In this work, for simplicity, the search is limited to molecules placed in general Wyckoff positions, since all targets from past CSP blind tests which are compatible with the ANI models are found in general positions~\cite{lommerse2000test, motherwell2002crystal, day2005third, day2009significant, bardwell2011towards, reilly2016report}. However, Genarris has the capacity to generate structures in special Wyckoff positions~\cite{li2018genarris,tom2020genarris}, which would be necessary in a more general search. To produce a single structure, a compatible space group is first chosen. The lattice vectors and the center of mass for a single molecule are randomly generated. Using the symmetry operations of a particular space group, the remaining molecules are placed within the cell. A proposed structure is valid if the vdW spheres of two atoms from different molecules do not significantly overlap. This is done by setting a parameter $s$ which restricts the distance $d$ between two atoms to be greater than $s(r_a + r_b)$, where $r_a$, $r_b$ are the vdW radii of atoms $a$ and $b$, respectively. We set $s=0.65$ by default. It should be noted that this value is less than the default value set in Genarris ($s=0.85$) \cite{tom2020genarris}. In this work, we generate much larger structure pools than those generated in the two studies of Genarris (Ref. \cite{li2018genarris}
and Ref. \cite{tom2020genarris}). We are able to use such large pools in our pipeline due to our reliance on machine-learned NNPs as opposed to quantum mechanical calculations. For larger molecules, we found that Genarris was unable to generate enough structures with the default value of $s$. Setting $s=0.65$ remedied this problem and allowed us to further explore the configuration space. For consistency, we chose to use this value across all of our runs. For more information about the crystal generation process, we refer the reader to the original work \cite{tom2020genarris}.

\subsubsection{Structural Relaxation and Property Computation}
Given the set of crystal structures, we compute desired properties and optionally perform structural relaxations. For the case of finding low-energy polymorphs, we seek to calculate the total energy of crystal structures. To rapidly compute total energies and perform structural relaxations, we chose to incorporate the ANI machine learning (ML) models with \cite{rezajooei2022neural} and without \cite{smith2018less, smith2019outsmarting, devereux2020extending} dispersion corrections, into our pipeline. The ANI models are trained on millions of organic molecules and are accurate across different domains. In addition, they have been shown to outperform many common force-fields in terms  of accuracy \cite{folmsbee2021assessing}. They are comprised of a set of artificial neural networks (ANNs) that use atomic environment descriptors as input vectors. Each ANN is specialized for a certain atom type and the output of an atomic ANN is the contribution to the total energy. The total energy is obtained by performing a summation of the atomic contributions. The final model is an ensemble model consisting individual ANI models trained on different folds of the dataset. The ensemble structure of ANI also allows one to calculate an uncertainty estimate for a given prediction according to the Query By Committee Approach (QBC)~\cite{smith2018less}.  The ANI models were trained on gas-phase molecules, however, the models can be used with periodic boundary conditions (PBCs) and can be directly applied to organic crystals~\cite{torchani}. Structural relaxations are done via the atomic simulation environment (ASE) package~\cite{larsen2017atomic}. 
It should be noted that our pipeline is not limited to the use of the ANI models and any ML model can be straightforwardly incorporated. 

\begin{figure}[bt]
     \centering
     \includegraphics[width=0.5\textwidth]{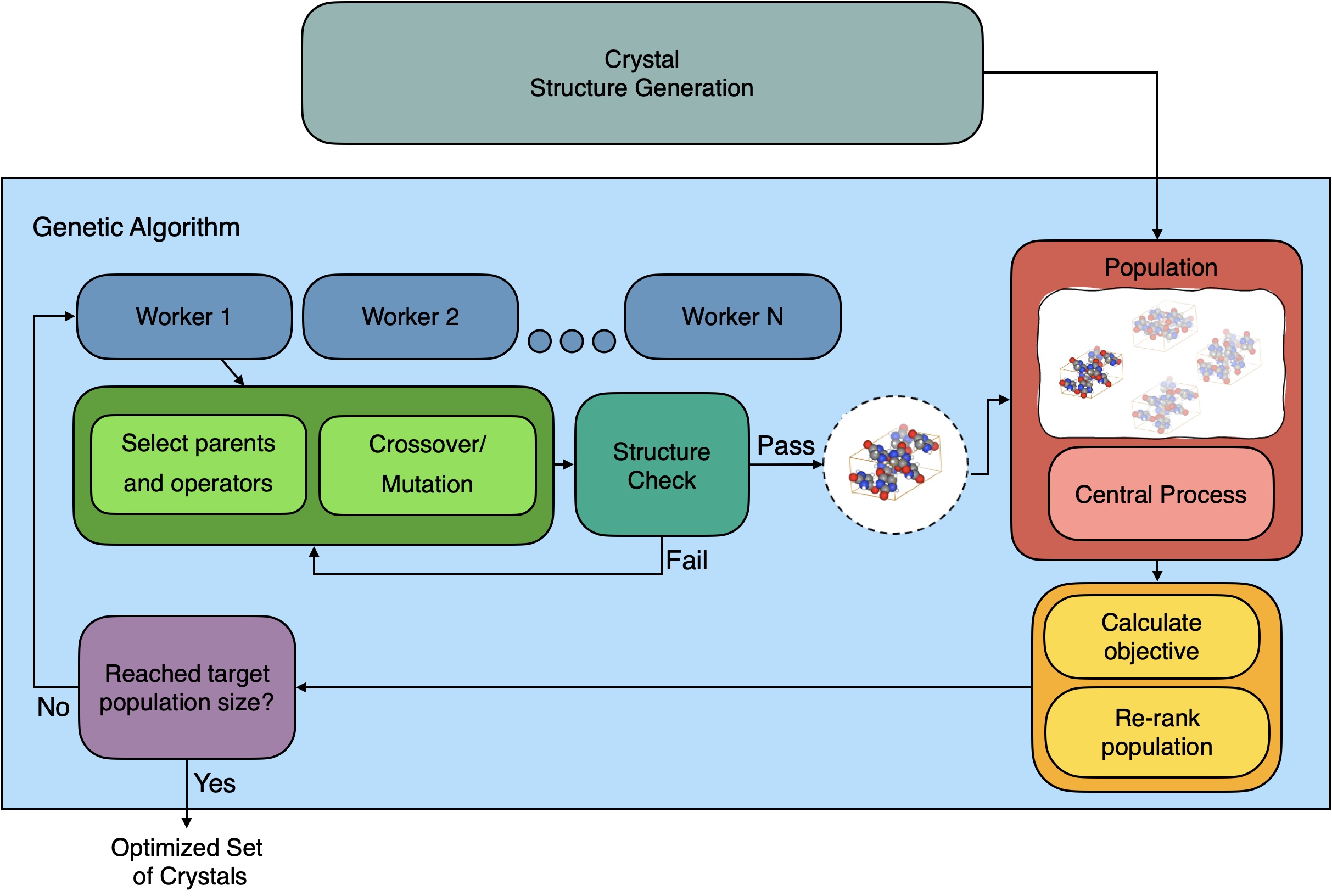}
     \caption{Visual representation of genetic algorithm. One central process handles population-level procedures -- calculating the objectives of structures in the population, re-ranking the population, and distributing it to the worker processes. Each worker process handles selection and breeding, and returns a valid structure to the central process.}
     \label{fig:ga_overview}
\end{figure}

\subsection{Genetic Algorithm}

Our GA implementation (seen in \cref{fig:ga_overview}) receives an initial population from random search, and outputs an optimized pool of structures according to an objective function. 
Our GA can be tailored to any objective function which can be computed for an individual molecular crystal, e.g. ANI energy, DFT energy, RMSD with respect to a target structure etc. Our GA utilizes an MPI approach with $n$ processes, where $n-1$ worker processes are responsible for selecting parents, applying breeding operators, optionally applying structural relaxation, making sure that the newly generated structure is physically viable, and not a duplicate of another structure in the population. Both structural relaxation and structure checking (making sure the structure is physically viable) are done in the same way as in \cref{subsec:random_gen}. The remaining process receives a newly generated structure from each of the worker processes, calculates the objective functions for each structure in the population, re-ranks the population, and distributes the latest population.

\subsubsection{Filtering and Clustering}
To create a refined pool of structures to form the initial population for the GA, we use filtering, and clustering using affinity propagation (AP)~\cite{frey2007clustering} on the pool of structures output by random search. Filtering is straightforward -- structures are ranked according to some property and the top structures are retained. In this paper, we only filter using the objective (ANI energy or  heavy-atom RMSD), however, one could filter on any property unrelated to the objective, i.e., volume, solubility, etc. Clustering via AP is performed with scikit-learn \cite{scikit-learn}. We use the relative coordinate descriptor (RCD) of each crystal as feature vectors for clustering. The RCD was introduced in Genarris~\cite{li2018genarris} and captures the relative orientation and positioning of molecules in a molecular crystal. Both RCD and the radial symmetry function (described in Ref.\cite{behler2007generalized}) can be used to generate feature vectors for clustering in Genarris~\cite{tom2020genarris}. Both were shown to yield similar performance. The Euclidean distance between two structures' RCD vectors is used to compute similarity. 
When clustering is used in our pipeline, a single round of AP is run. The top structures according to the target property are selected from each cluster to satisfy the required number of structures.

\subsubsection{Selection}

One or two parent structures may be selected to produce offspring depending on which breeding operator is selected. Each selection strategy depends on the fitness of individuals in the population. Each time the objective function is evaluated, objectives are normalized to yield a probability distribution over the population. We use the same objective function as GAtor \cite{curtis2018gator}. For structure $i$ the objective function is
\begin{equation}
    f_i = \frac{\epsilon_i}{\sum_{i=1}^{N}\epsilon_i}
\end{equation}
where
\begin{equation}
    \epsilon_i = \frac{E_{\text{max}} - E_i}{E_{\text{max}} - E_\text{min}}.
\end{equation}
Where $E_i$ is the target property of structure $i$, and $E_{\text{max}}, E_{\text{min}}$ are the maximum and minimum target property found across the entire current population. In the present study, $E_{i}$ is limited to the energy or RMSD of structure $i$ with respect to a target structure.

Various selection strategies are included in our implementation. Both roulette wheel and tournament selection as implemented in GAtor~\cite{curtis2018gator} are included. For brevity, we quickly summarize these selection schemes. For roulette wheel selection, the objective values of candidates are normalized to yield a probability distribution over the population and then used to directly sample from the population and select parents. For tournament selection, a small pool of structures is first randomly chosen. Afterwards, the pool is ranked according to the objective, and the top-ranked structures are selected. 

We also include two new selection strategies: \texttt{top} and \texttt{uniform}. When selecting using \texttt{top}, the $k$ structures with the highest objective are set aside (where $k$ is a parameter set by the user), and parents are randomly sampled from this smaller pool. When selecting using \texttt{uniform}, all structures have an equal probability of being selected. Lastly, we include the option to use any of the four selection strategies at random, which can be chosen by setting the selection strategy to \texttt{all} in GAmuza. The authors of GAtor \cite{curtis2018gator} recommend trying different parameters within the GA and collecting the results of different GA runs. Setting the selection rule to \texttt{all} allows us to achieve a similar effect.

\subsubsection{Breeding Operators}

Breeding operators are grouped into either crossover or mutation operations. In GAmuza, the likelihood of performing a crossover operation can be set. However, by default, the likelihood of performing a crossover operation is the same as a mutation operation. Crossover operations blend the genes of two parent structures, while a mutation operation modifies a single parent structure to produce a child. Our GA includes all of the original breeding operators implemented in GAtor~\cite{curtis2018gator}, plus two newly developed symmetry-preserving mutation operations.

\textbf{Crossover:} We include both original crossover operations in GAtor: standard and symmetric crossover. Since our implementation may have multiple molecular conformers, we note that crossover operations are restricted to structures with identical conformers. Both crossover operations are tailored to molecular crystals. During crossover, each structure is represented by its Niggli reduced, standardized cell, and the centre of masses and orientations of the molecules in the structure. These three properties (cell, positions, and orientations) may be blended with random fractions, or directly inherited from one of the parents. Symmetric crossover ensures that the child structure inherits the space group of one of its parents, while standard does not.

\begin{figure}[bt]
     \centering
     \includegraphics[width=0.5\textwidth]{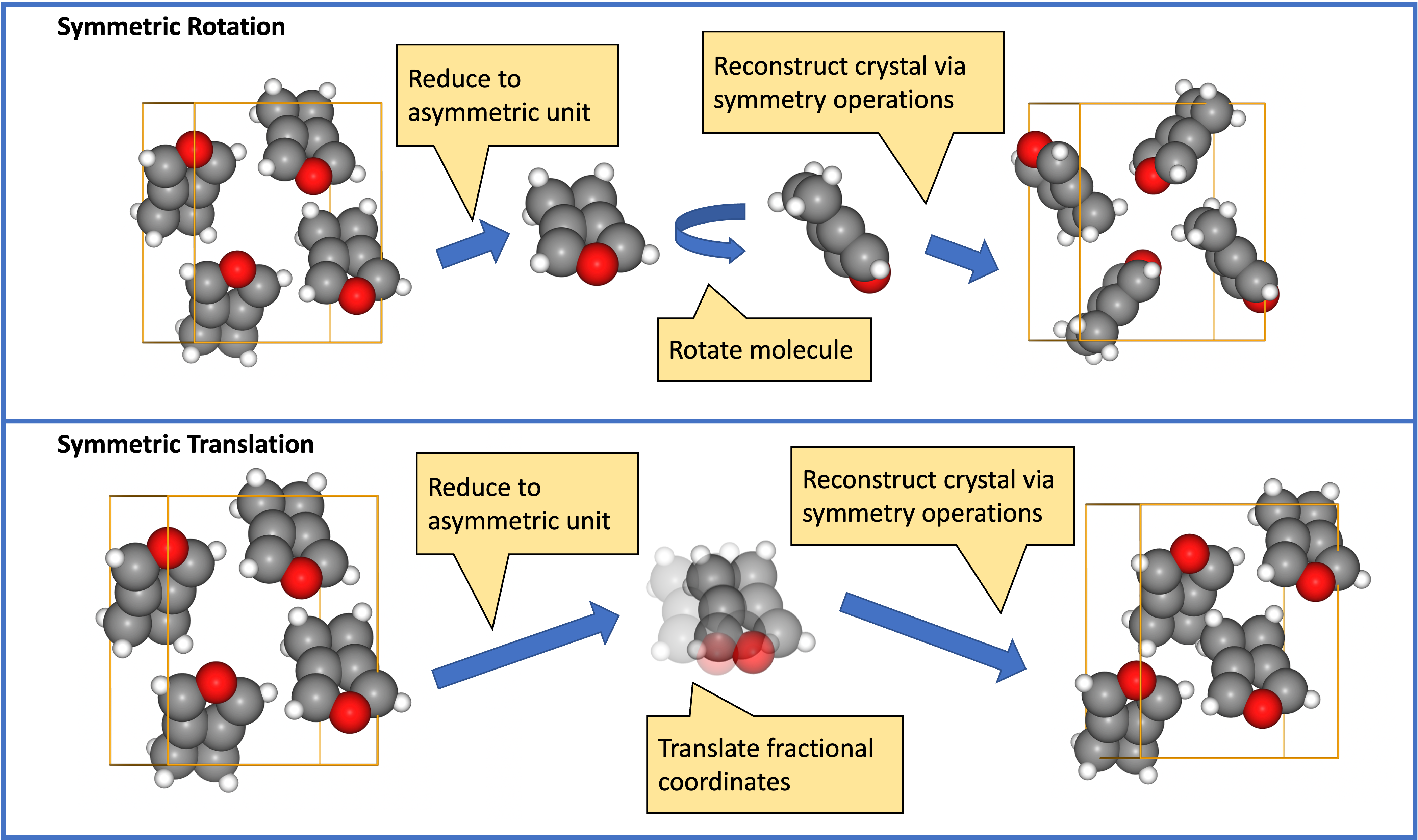}
     \caption{New operations introduced in GAmuza. Both symmetric rotation and symmetric translation allow one to rotate and translate molecules within the crystal while conserving the space group of the parent structure. This is achieved by extracting the asymmetric unit from the crystal, applying operations to the individual molecule, and regenerating the structure via the symmetry operations of the space group.}
     \label{fig:sym_ops}
\end{figure}

\textbf{Mutation:} We include all of the original mutation operations in GAtor~\cite{curtis2018gator}, which we quickly summarize. These include two operations for translation: random translation, and frame translation. During random translation, a random selection of molecules in the unit cell are translated by a random vector. When using frame translation, each molecule is translated in a random direction in the basis of its inertial reference frame, constructed via each molecule's principal axes of rotation. GAtor also includes two operations for rotating the molecule: random rotation and frame rotation. In a similar spirit to the translation operations, random rotation applies a random rotation to a random subset of molecules in the unit cell, while frame rotation applies the same rotation to each molecule in its inertial frame of reference. GAtor also includes three permutation mutations: permute molecule, permute-rotate, and permute-reflect. In permute molecule, two molecules' centres of mass in the unit cell are randomly exchanged. During permute-rotate, two molecules' centres of mass are exchanged, followed by a random rotation in each molecule's inertial frame of reference. When permute-reflect is applied, once again the centre of mass of two molecules in the unit cell is exchanged, followed by a random reflection in either the $x$, $y$, or $z$-axis. Lastly, GAtor contains four strain mutations: random strain, angle strain, volume strain, and symmetric strain. Each strain mutation applies a stretch and/or shear to the lattice. The fractional coordinates of the molecules are moved accordingly.

In addition to all of the breeding operators from GAtor, we have added two additional symmetry-preserving mutation operations which allow one to translate and rotate the molecules in the crystal without breaking symmetry. This is accomplished by first choosing a representative molecule in the crystal. Once this molecule is chosen, the symmetry operations of the crystal system's space group are calculated.

A rotation or translation is applied to the representative molecule, and the symmetry operations are used to re-generate the entire crystal structure. This is in alignment with how symmetric crossover is implemented in GAtor. A graphic depicting these two new operations is included in \cref{fig:sym_ops}.

\subsubsection{Evolutionary Niching}

We have included the option to perform evolutionary niching in GAmuza as described in \cite{curtis2018evolutionary}. Evolutionary Niching modifies the objective function such that structures with less structural similarity to other structures in the population are more likely to be selected. Doing so promotes more diverse structures in the population. This is achieved by running AP with RCD each time a new structure is added to the population. Each crystal's objective is then divided by the number of structures that are present in its cluster. The major computational bottleneck when clustering is to calculate the pairwise RCD distances. Therefore, the pairwise distances are stored in memory, and only the row and column corresponding to the newly added structure need to be calculated, which, in combination with running AP again, takes only a few seconds.

\subsection{Platform Availability}

The pipeline is available on QEMIST Cloud, a cloud-based simulation platform that includes various computational chemistry techniques (i.e. DFT and incremental full configuration interaction~\cite{zimmerman2017incremental}). Any of these various simulation techniques can be used to compute properties and perform structural minimization of the generated crystals. In addition, the pipeline is not limited to total energy but can be adjusted to accommodate any property which one can compute for a single crystal structure.

\section{Results}
\label{results}

\begin{figure}
    \centering
    \includegraphics[width=\linewidth]{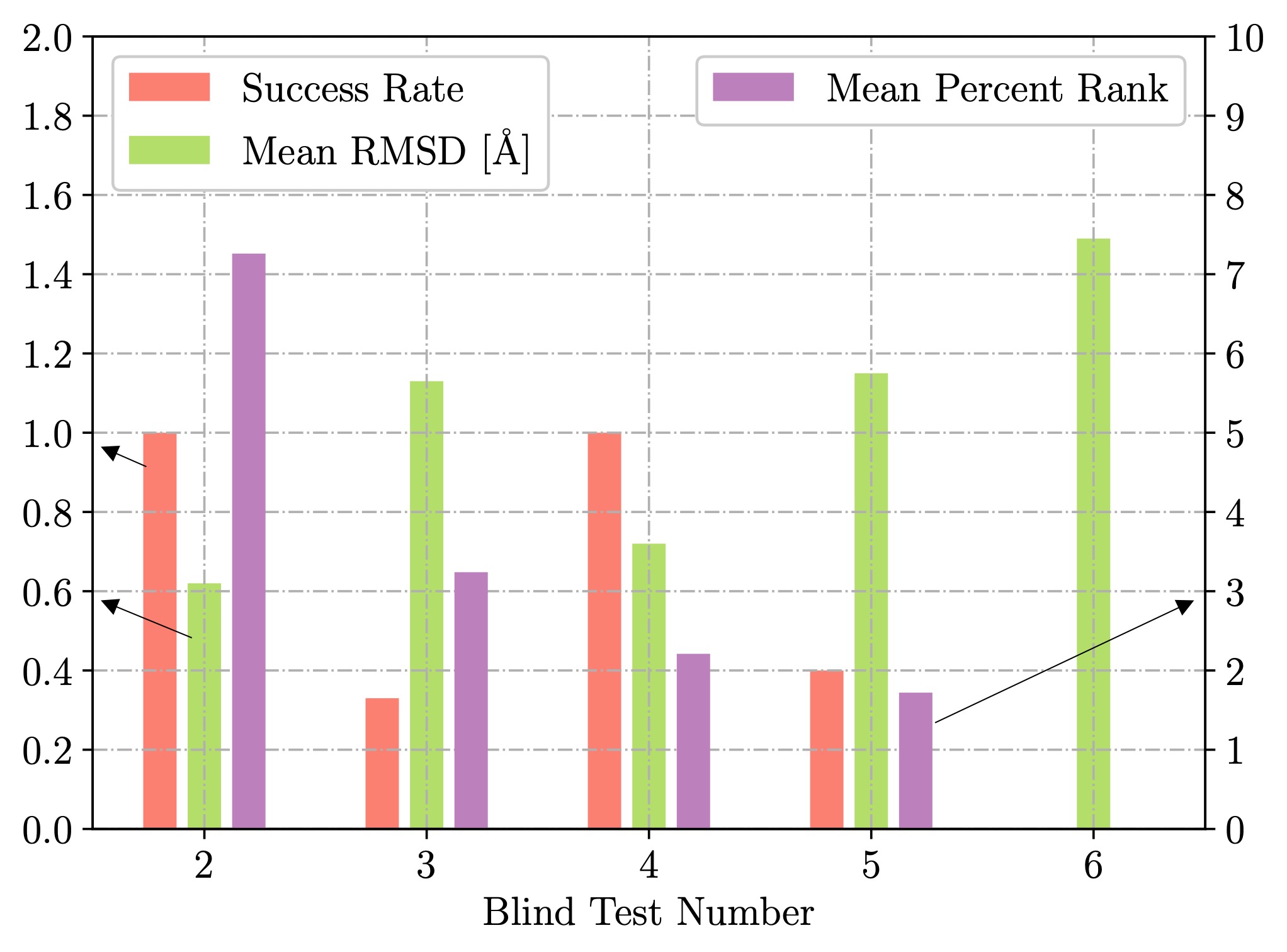}
    \caption{The success rates, mean root mean-squared deviation (RMSD), and mean percent ranking (for successful matches) when using random crystal generation with the ANI2x model. Arrows indicate the correct axes.}
    \label{results_1}
\end{figure}

\subsection{Random Structure Generation and Ranking with ANI2x}\label{subsec:random_gen_results}
We assess the performance of the ANI2x model when applied to ranking organic crystals. These results can be seen in \cref{results_1} and \cref{ranking_table} in \cref{si}. To do so, we consider all blind test targets with atom types that are supported by the ANI2x model (C, N, O, H, F, Cl, and S), making  for 21 molecules. For target  xxiii from the sixth blind test, we report results for all 5 known stable polymorphs. Blind test submissions are evaluated on whether they can produce a matching structure for a given ground-truth structure consisting of an experimentally derived polymorph~\cite{lommerse2000firstblindtest}. A proprietary matching algorithm which is part of the closed source CSD python API \cite{sanschagrin2017using} is used to determine whether a given structure is a match. Roughly speaking, a structure is considered a match if it within 0.8 \AA\; RMSD from the ground-truth structure \cite{lommerse2000firstblindtest,motherwell2002secondblindtest,day2005thirdblindtest,day2009fourthblindtest,bardwell2011fifthblindtest,reilly2016sixthblindtest}, which is the definition we adopt in this manuscript.

To calculate RMSD, we use our own matching technique, employing pymatgen's \texttt{StructureMatcher}~\cite{ong2013pymatgen} to compare generated structures with the experimentally derived polymorph. \texttt{StructureMatcher} is designed to operate on similar crystal structures, and has been used for crystal comparison in a number of other works~\cite{hu2023deep,zimmermann2020local,chen2019investigation,wei2022tcsp}, including both GAtor~\cite{curtis2018gator}, and Genarris~\cite{li2018genarris}. Using pymatgen's \texttt{StructureMatcher}, a generated structure is put into a similar basis to the reference structure without changing any of the inter-atomic distances. Then, we find an optimal translation and rotation of the mapped structure, such that the RMSD between corresponding heavy atoms is minimized. We use \texttt{StructureMatcher} with \texttt{ltol=0.6}, \texttt{stol=0.9}, and \texttt{angle\_tol=15}. These values are much larger than the default values due to the potential large discrepancy between generated structures and the experimentally derived polymorph. We then calculate the RMSD using only the heavy atoms. 

When performing random search, for each molecule we generate structures that have 2,4, and 8 number of molecules per cell ($N_{mpc}$), as is standard when submitting to the blind tests~\cite{lommerse2000firstblindtest,motherwell2002secondblindtest,day2005thirdblindtest,day2009fourthblindtest,bardwell2011fifthblindtest,reilly2016sixthblindtest}. For each setting of $N_{mpc}$, we generate $10^x$ number of structures per conformer where $x\in \{2, 3, 4, 5, 6\}$. We start with $x=2$ and increase the population size until either a match is found or the required compute time to generate the desired pool size exceeds 5000 CPU hours. To account for the effect of the random seed in our pipeline, for each experiment (i.e. setting of $N_{mpc}$ and population size), we run 10 independent runs, each with a different random seed. For each generated crystal structure, we compute its ANI energy and RMSD with respect to the ground truth structure. For each experiment, we record how often a match was found across the 10 runs, the lowest ranking match when combining all structures across the 10 runs (total rank), the percentage of structures which were ranked lower than the lowest ranked match (\% rank), the best ranking achieved by a match when analyzing each of the independent runs individually (best rank), the minimum RMSD achieved in the experiment, and the CPU hours to run that experiment.

Through this procedure, we were able to produce matches for 10/21 molecules. Furthermore, we found a match for at least one molecule from each of the first five blind tests. However, we did not find success for the 6th blind test with random generation. This is not surprising, as the average success rate for teams employing random generation/search had a 17\% success rate\cite{reilly2016sixthblindtest}. We note the number of molecular conformers is equal to 1 for most of the successful cases whereas the number of molecular conformers is $>1$ for the failed cases. For candidates with $N$ molecular conformers, the size of the chemical space that must be explored increases by a factor of $N.$ Therefore, with enough computation (i.e. $>$ 5000 CPU hours) one will find success with random search, even for cases with multiple molecular conformers. 

For 1/10 cases, ANI2x ranked a match within the top 100 structures, and for 2/10 cases, a match was ranked within the top 500. This is important to note since the maximum number of structures that could be submitted to the 6th blind test was 100 and 100-500 for the 7th blind test, depending on the molecule.

Of the cases where matches were found, matches were ranked on average in the top 4.7\% by ANI2x. This average is driven up by two targets, i and ii, where ANI2x exhibits significantly worse ranking power than it does for other targets, ranking each top match just outside of the first 15\% of generated structures. For the 8 other structures, each top match is ranked within the top 4\% of generated structures. For 3 of these (iv, vi, vii), each top match is ranked within the top 1\%.  

Overall, ANI2x is currently better suited as a filtering tool before using some higher accuracy method for final ranking. Using ANI2x as a scoring method allows you to filter out $>95\%$ of generated structures. This is valuable in reducing the amount of time that is spent ranking lower-quality samples. However, when one needs to submit a list of 100/500 structures from potentially millions of generated structures, ANI2x is insufficient on its own.

\subsection{Comparing the Ranking ability of other ANI models}
\label{comparing}

\begin{figure}
    \centering
    \includegraphics[width=\linewidth]{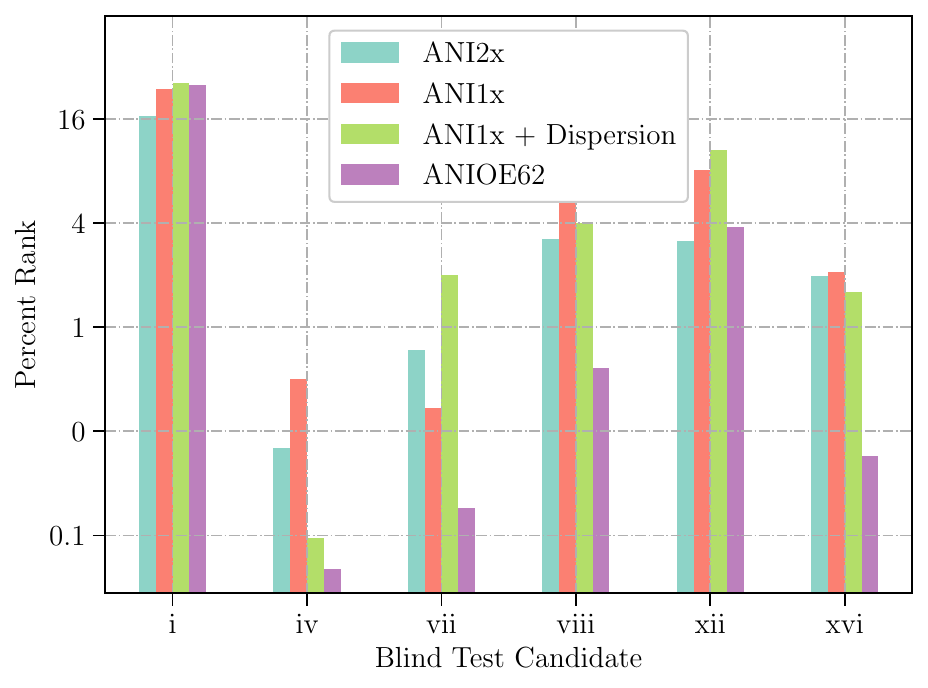}
    \caption{Comparison of the ranking abilities of various ANI models for different past blind test candidate molecules. The natural logarithm was taken for visualization purposes and 1 was added to the percent rank to avoid negative values. }
    \label{result_2}
\end{figure}
We now consider different ANI models to re-rank the matches found in \cref{subsec:random_gen_results}. We consider ANI1x and ANI2x from \texttt{torchani} \cite{torchani}, ANI1x with dispersion from \texttt{torchanipbe0} \cite{rezajooei2022neural}, and ANIOE62 -- a model we introduce. The model name is derived from the OE62 dataset, which is a collection of 61,489 molecules extracted from organic crystals in the Cambridge Structural Database (CSD)~\cite{stuke2020atomic,allen2002cambridge}. Each molecule in the OE62 dataset was relaxed via DFT using the PBE exchange-correlation functional \cite{perdew1996generalized}, including  Tkatchenko-Scheffler (TS) vdW corrections \cite{tkatchenko2009accurate}. For each relaxed geometry, the total energies and orbital eigenvalues with both the PBE and PBE0 functionals~\cite{adamo1999toward} are provided. 

To construct the dataset for transfer learning with ANI2x~\cite{devereux2020extending}, we first filtered out geometries of molecules containing atoms not compatible with ANI2x (H, C, N, O, F, Cl, S). This resulted in 49,514 configurations for training. We initialized ANI2x with parameters from \texttt{torchani} and then re-trained the model to predict the total energies at the PBE0 level. Each model in the ensemble received a random sample of 80\% of structures from the dataset and was trained for 300 epochs with the Adam optimizer~\cite{kingma2014adam} on the mean squared error loss, with a batch size of 1024 molecules. The Adam optimizer was used with an initial learning rate of $10^{-4}$, and the learning rate was halved if the training loss did not decrease for 50 epochs. Each model was trained on a single NVIDIA V100 GPU instance (on Amazon Web Services). Training and validation curves can be found in \cref{si}.

Based on \cref{result_2}, and \cref{different_ani_models} in \cref{si}, we find that ANIOE62 tends to offer the most accurate ranking of a matched structure, eliciting the best ranking for 4/6 targets. Somewhat surprisingly, ANI2x offers the best ranking for the two other targets. To further understand why ANI2x ranks some targets much better than others, we use the uncertainty estimate described in Ref.\cite{smith2018less}, which consists of the standard deviation of the ensemble model divided by the square root of the number of atoms. For the cases where ANI2x had a percent rank $\leq$ 5\%, we found that the uncertainty estimate was 1.8 times lower than for case where the percent rank was $>$ 5\%. This indicates that ANI2x has not been trained in this area of chemical space and we expect that retraining would improve its ranking ability. We find that dispersion corrections can at times, significantly change the ranking of ANI1x, but it does not seem to always help, improving the ranking of ANI1x for half of the targets. This is surprising, as intermolecular interactions within the crystals come from vdW forces and hydrogen bonding. It is well understood that plane-wave-based DFT methods can accurately describe the energetics of organic crystals~\cite{reilly2013understanding, marom2013many, kronik2014understanding, moellmann2014dft, reilly2016sixthblindtest,hoja2018first, price2023xdm}. In addition, the inclusion of vibrational effects also plays an important role in accurate polymorph ranking \cite{reilly2016sixthblindtest}. Part of our future work will be including vibrational corrections with an ML model, to maintain the speed of our pipeline. Likely, the final missing ingredients to improve the ranking capabilities of the ML model come from differences in the electronic structure between isolated and crystallized molecules and the inclusion of vibrational effects. We hypothesize that an organic crystal dataset based on plane-wave-based DFT calculations (using the PBE0 functional and many-body dispersion corrections) along with vibrational free energies computed via ML would be the best way forward. We plan to explore this in our future work.

\subsection{Validation of the Genetic Algorithm}\label{subsec:valid_ga}

\begin{figure*}
    \centering
    \includegraphics[width=\linewidth]{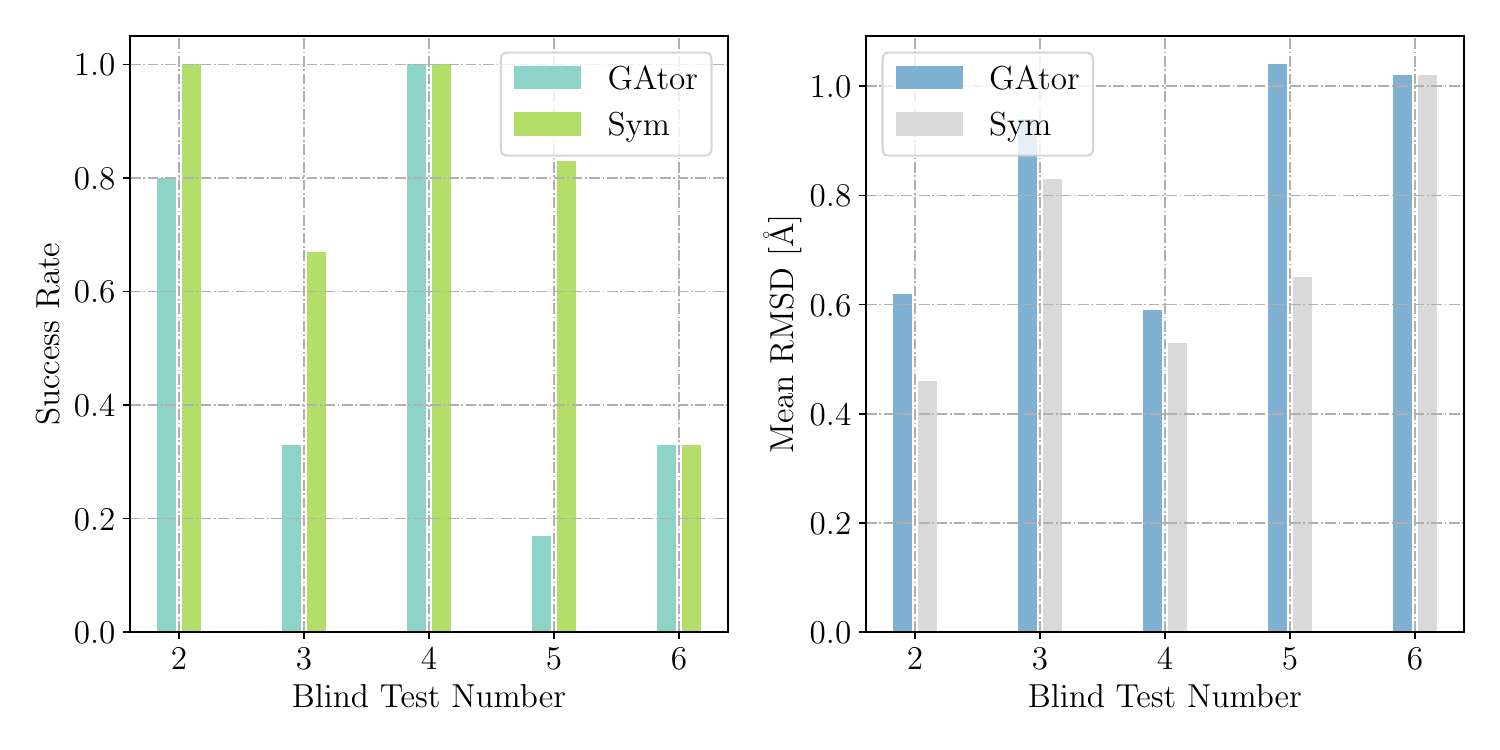}
    \caption{The success rate (left) and the mean root mean-squared deviations (RMSDs) (right) when using the RMSD objective function for the past blind tests using the GAtor or symmetric (Sym) setting within GAmuza.}
    \label{result_3}
\end{figure*}
In this section, we analyze the sampling power of the GA in GAmuza. In order to isolate its power from the ranking ability of the ANI NNPs, we equip the GA with an idealized energy function. The results are reported in \cref{result_3} and \cref{table:ga_results} of \cref{si}. We focus on optimizing the RMSDs of crystal structures with respect to the experimental structures. This allows us to determine whether the operations in the GA are sufficient for producing structures similar to known stable polymorphs when equipped with a scoring method that has these stable polymorphs as global minima in its energy function. 

We consider all 21 molecules from \cref{subsec:random_gen}. However, due to computational constraints, we only consider a single stable polymorph of target xxiii from the sixth blind test (xxiii-A). Since we are now using RMSD to guide the GA, running each stable polymorph would require 5 independent runs, unlike the previous section where the results from a single run guided by energy could be compared to all 5 polymorphs. For each target, we run our entire pipeline, beginning with random search, followed by the GA. We follow a similar procedure as in \cref{subsec:random_gen_results}. Each time we run our pipeline on a new target, we experiment with increasing the initial pool size until a match is found, or the computational cost becomes $>$ 5,000 CPU hours. Once again, this is achieved by generating $10^x$ structures per conformer for $x \in \{2,3,4,5,6\}$. As in \cref{subsec:random_gen_results}, each run is repeated 10 times to account for randomness in the pipeline.

We run the GA with two different configurations. In the \texttt{GAtor} configuration, we run the GA with all of the original breeding operators in GAtor \cite{curtis2018evolutionary}. In the \texttt{symmetric} configuration, only breeding operators which preserve the space group of the parents of the child structure are considered. In this configuration, symmetric crossover is used as the sole crossover operation. Mutations include all of the strains from GAtor, as well as our implemented symmetric translation and rotation. Both configurations use \texttt{all} as their selection rule. 

For each run, we first filter the initial pool down to 5,000 structures, followed by clustering to reduce the initial pool fed into the GA to 100 structures. If the initial pool is less than 5000, we skip filtering, and if it is less than 100, we skip clustering too. We use the GA to generate 900 new structures, bringing the final population to 1000. A filter size of 5000 was chosen, as it takes between 5 and 10 minutes on 32 cores to calculate the pairwise RCD distances for this many structures. For $n$ structures, calculating pairwise distances is $O(n^2)$, and hence, doubling the pool size would quadruple the time taken to calculate pairwise distances.

For each target, RMSDs are calculated with our matching technique, specified in \cref{subsec:random_gen_results}. At the end of each run of the GA, we tally the number of runs which have produced a match -- a structure with RMSD $<0.8~\text{\AA}$ relative to the experimental structure. We report the minimum RMSD structure produced by the pipeline, the rank of the lowest-ranked matched structure, and the computational time required to run our pipeline to completion 10 times with the reported pool size. 

We find a $\approx60\%$ increase in the number of matches (16/21 targets matched) when using this idealized objective function with the GA versus random structure generation alone. Notably, the GA is able to discover two targets from the most recent 6th blind test which were not discovered by random generation: xxii and xxiii-A. In addition, the GA allows one to greatly reduce the number of crystals generated in comparison with random search, requiring anywhere from 10 to 1,000 times fewer structures in the initial pool. Lastly, it can be seen that the GA with the \texttt{symmetric} configuration generally outperforms the \texttt{GAtor} configuration, finding a match 5 times when \texttt{GAtor} does not.

\subsection{Performance of Entire Pipeline with ANI2x and ANIOE62}

\begin{figure*}
    \centering
    \includegraphics[width=\linewidth]{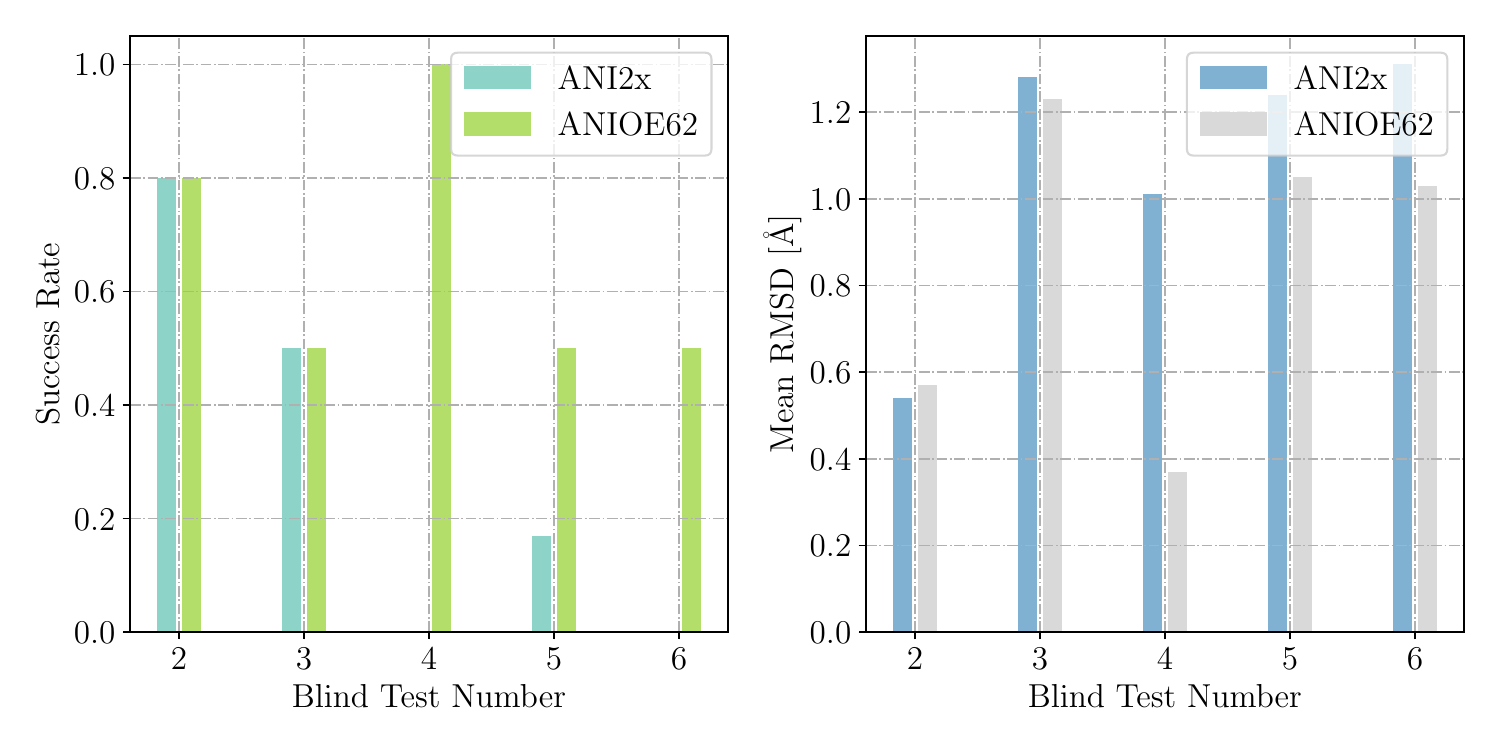}
    \caption{The success rate (left) and the mean root mean-squared deviations (RMSDs) (right) when using either ANI2x or ANIOE62 as the objective function for the past blind tests with the Symmetric setting within GAmuza.}
    \label{result_4}
\end{figure*}

\begin{figure}[bht]
     \centering
     % \hspace*{-1  cm} 
     \includegraphics[width=\linewidth]{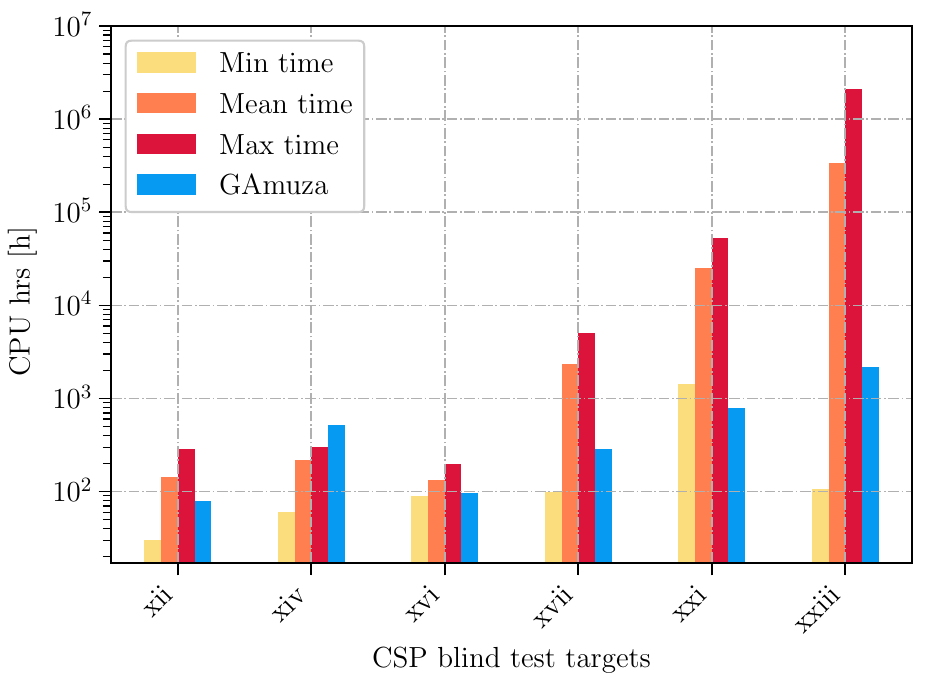}
     \caption{The time taken for GAmuza to generate various targets from the 4th, 5th, and 6th blind test when compared to other submissions. The y-axis is plotted on a logarithmic scale. The submissions with the least and highest run times are plotted, as well as the average run times for all successful submissions.}
     \label{fig:csp_times}
\end{figure}

\begin{figure*}[bht]
    \centering
    \includegraphics[width=\linewidth]{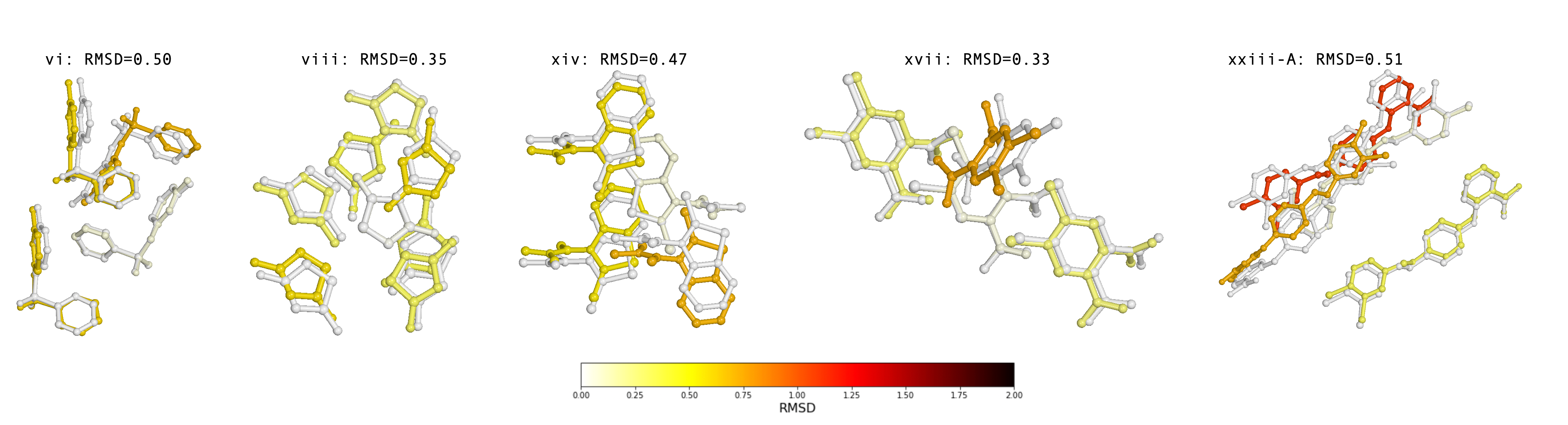}
    \caption{Results of various targets attained when running our pipeline from start to finish. The experimental structure is visualized in grey. The structure generated with GAmuza is fitted to the experimental structure, and coloured to show the RMSD of individual molecules from the experimentally derived structure. Only molecules within the unit cell are visualized. Visualizations are done with the Molecular Crystal Simulation Environment~\cite{MCSE}.}
    \label{fig:matches}
\end{figure*}

Finally, we run our entire pipeline from start to finish using energy calculated with the ANI NNPs as the objective. The results can be seen visually in \cref{result_4} and in \cref{table:full_results} of \cref{si}. This allows us to assess the current performance of our pipeline. Based on the results of \cref{comparing} and \cref{subsec:random_gen_results}, we do not believe that the current ranking ability of ANI is sufficient for situations where one requires an extremely accurate final ranking of crystals (such as when submitting to the blind tests). On top of assessing the current state of our pipeline, these results allow us to understand the degree of improvement that is required to have a successful CSP pipeline using NNPs.

We limit results to targets for which we were able to successfully produce a match using RMSD as the objective. For target xxiii, as in \cref{subsec:random_gen}, we report the results for all 5 stable polymorphs.
For each target, we try various initial population sizes. We begin by setting the initial population size to that which was used to yield a match when using RMSD as the objective (from~\cref{table:ga_results}). Since the purpose of the GA is to reduce the population sizes required by random search via a more efficient exploration of the objective landscape, we limit the initial population size to $10\times$ less than what was used for random search in Section~\ref{subsec:random_gen_results}. We use the \texttt{symmetric} configuration, as this was shown to produce the best results in the previous section. With the ANI models guiding the search, we also can relax structures before adding them to the population. When relaxing a structure, we use the BFGS algorithm \cite{fletcher2013practical} where the magnitude of the maximum force must be less than 0.1 eV / \AA. We make sure to conserve the symmetry of the space group using the \texttt{FixSymmetry} constraint available when running the relaxations with ASE~\cite{larsen2017atomic}, allowing us to further conserve symmetry throughout our pipeline. We use both the original ANI2x and ANIOE62 to guide the search, as these are the only two models that can be run on every target, and were top 2 performing models in terms of ranking performance, as show in Section~\ref{comparing}.

In total, we can find matches for 12/16 molecules by running our full pipeline. We find 6 of those matches when ANI2x is guiding the search, and 11 matches when ANIOE62 is used to guide the search. ANI2x found a match only once (target vi) when ANIOE62 failed, while ANIOE62 found matches six times (targets ii, xii, xiv, xvi, xxi, and xxiii) when ANI2x failed. For target xxiii, ANIOE62 recovered 2/5 stable polymorphs (xxiii-A and xxiii-B). This further validates that this newly trained model exhibits better ranking ability for organic crystals. For 4/12 targets, we found a match in the top 100 ranked structures, and for 7/12 targets we found a match in the top 500 structures. Once again, this is important to note since the maximum number of structures that could be submitted to the 6th blind test was 100 and 100-500 for the 7th blind test, depending on the target. Matches found by our pipeline are visualized in Figure~\ref{fig:matches}.

In comparison to random generation, the GA can find 2 more matches (20\% increase), successfully finds a match in the top 100 structures for 2 more targets (100\% increase), and successfully finds a match in the top 500 structures for 5 more structures (250\% increase). This was also shown to be the case in Ref. \cite{curtis2018evolutionary}, where GAtor found success when random generation did not. Since the GA is making incremental improvements to current structures rather than starting from scratch (i.e. random generation), it can surpass the success rate of random generation alone. The GA requires 10 to 100 times fewer structures in the initial pool relative to random generation, once again exhibiting that our GA can effectively explore the potential energy landscape. For simpler targets, the GA is often slower than random search. However, the power of the GA is established when comparing more complex structures. To generate a match for xxiii-A with random search would have taken $>10,000$ CPU hours, whereas with the GA we find a match in under 3000 CPU hours.

Lastly, we compare the run time of our method to successful submissions in the 3 latest blind tests. The run times can be seen graphically in~\cref{fig:csp_times}. For 4/6 targets GAmuza achieves a faster run time than the mean successful run time, but a slower run time than the fastest successful run time. For one target (xxi), GAmuza is faster than the fastest successful method, and for one target (xiv), GAmuza is slower than the slowest-running successful method. Generally, GAmuza approaches speeds similar to methods employing force-fields for ranking, but is orders of magnitude faster than methods using \textit{ab initio} methods. This exhibits the power of using ML for ranking in CSP. ML is almost as fast as force-field-based methods, yet, as more and more data specific to organic crystal systems is collected, the more accurate the ranking method becomes, eventually approaching the accuracy achieved by quantum chemistry methods.

\section{Conclusion}
\label{conclusion}

In this work, we presented an organic CSP pipeline utilizing NNPs for fast, accurate ranking and relaxations of crystal structures, integrated into a GA for effective sampling of the search space. Many successful modern CSP approaches utilize DFT to achieve an accurate ranking of generated crystal structures \cite{reilly2016sixthblindtest}. NNPs are an efficient alternative, which can rival the accuracy of the underlying training data. Our pipeline pairs a family of ANI models to rank and relax crystal candidates, with a powerful pipeline, capable of proposing up to millions of candidate crystal structures and further refining these candidates via a genetic algorithm. We build on previous work, namely Genarris \cite{tom2020genarris} and GAtor \cite{curtis2018gator}, and include new symmetry-preserving operations that show improved performance compared to symmetry-breaking operations. We exhibit the potential of this pipeline, which can generate several crystal candidates from the 6 CSP blind tests via random generation alone, and even more candidates when GAmuza is paired with an accurate ranking objective function. We study the current power of this pipeline, pairing it with an NNP which has been transfer-learned to improve its predictive ability on organic crystals. Our pipeline is orders of magnitude faster than frameworks employing DFT. To the best of our knowledge, this is the first generally applicable CSP pipeline which makes use of a potential that approaches the speed of the fastest methods available and is continually improvable as more and more data is collected. 

This work also shows, that although the ANI NNPs are within chemical accuracy of DFT for energies evaluated on gas-phase organic molecules \cite{smith2018less}, the same cannot be said for organic crystals. We hypothesize that an organic crystal dataset would allow one to train an NNP capable of approaching and eventually surpassing the performance of CSP pipelines employing DFT. This is part of our future work.

\bibliography{refs}

\appendix

\section{Supplemental Information}
\label{si}

\begin{table*}[ht]
    \centering
    \scriptsize
    \begin{tabular}{c|c|c|c|c|c|c|c|c}
         Structure & $N_{c}$ & $N_s$ & Successful Runs & Total Rank & \% Rank & Best Rank & Min. RMSD [\AA] & Cost [CPU hr]\\\hline
i & 1 & 300 & 3 & 501 & 16.70 & 46 & 0.68 & 2.88 \\
ii & 1 & 30,000 & 2 & 55932 & 18.64 & 5743 & 0.74 & 11.14 \\
iv & 1 & 3,000 & 3 & 61 & 0.20 & 9 & 0.53 & 10.35 \\
vi & 5 & 1,500,000 & 4 & 917 & 0.01 & 94 & 0.63 & 1601.49 \\
vii & 1 & 3,000 & 6 & 223 & 0.74 & 25 & 0.52 & 7.77 \\\hline
viii & 1 & 30,000 & 5 & 9717 & $3.24$& 976 & 0.62 & 15.18 \\
x & 1 & 3,000,000 & 0 & - & - & - & 1.42 &  5173.33 \\
xi & 1 & 3,000,000 & 0 & - & - & - & 1.35 & 896.96 \\\hline
xii & 1 & 30,000 & 1 & 9523 & $3.17$ & 999 & 0.78 & 12.74 \\
xiv & 2 &  60,000 & 1 & 7493 & $1.25$ & 674 & 0.66 & 104.85 \\\hline
xvi & 1 & 30,000 & 1 & 5896 & $1.97$& 644 & 0.58 & 13.09 \\
xvii & 1 & 30,000 & 3 & 4442 & $1.48$& 117 & 0.69 & 29.08 \\
xviii & 2 & 600,000 & 0 & - & - & - & 1.44 &  497.49 \\
xix & 60 & 1,800,000 & 0 & - & - & - & 1.11 & 2420.00 \\
xx & 10 & 300,000 & 0 & - & - & - & 2.19 & 1097.89 \\
xxi & 10 &  300,000 & 0 & - & - & - & 0.90 &  238.54 \\\hline
xxii & 2 &  6,000,000 & 0 & - & - & - & 0.88 &  4050.67 \\
xxiii-A & 9 & 270,000 & 0 & - & - & - & 1.38 &  1062.89 \\
xxiii-B & 9 & 270,000 & 0 & - & - & - & 2.00 &  1062.89 \\
xxiii-C & 9 & 270,000 & 0 & - & - & - & 2.15 &  1062.89 \\
xxiii-D & 9 & 270,000 & 0 & - & - & - & 3.41 &  1062.89 \\
xxiii-E & 9 & 270,000 & 0 & - & - & - & - &  1062.89 \\
xxiv & 120 & 3,600,000 & 0 & - & - & - & 1.32 &  4116.00 \\
xxv & 150 & 450,000 & 0 & - & - & - & 1.89 &  1468.67 \\
xxvi & 3 & 900,000 & 0 & - & - & - & 1.98 &  3986.09 \\

    \end{tabular}
    \caption{Ranking ability of ANI2x for various structures. $N_s$ is the number of randomly generated structures. $N_c$ is the number of molecular conformers considered. Successful Runs is defined to be the number of runs in which a match was generated (RMSD $<$ 0.8 \AA) over 10 trials with different random seeds. Total Rank and \% Rank are calculated correspond to the ranking of matches when compared to structures from all 10 runs. Best Rank is the lowest-ranked match in any of the individual 10 trials.
    Min. RMSD is the minimum root mean-squared deviation found with respect to the experimental structure.}
    \label{ranking_table}
\end{table*}

\begin{table*}[ht]
    \centering
    \begin{tabular}{c|c|c|c|c|c}
         Structure & $N_s$ & Model & Total Rank & \% Rank & Best Ranking \\\hline
         \multirow{4}{*}{i} & \multirow{4}{*}{300}  
         & \textbf{ANI2x} & \textbf{501} & \textbf{16.7} & \textbf{46} \\
         & & ANI1x & 712 & 23.7 & 68 \\
         & & ANI1x + Dispersion & 770 & 25.7 & 80 \\
         & & ANIOE62 & 757 & 25.2 & 72  \\
         \hline
         \multirow{4}{*}{iv} & \multirow{4}{*}{3000}
         & ANI2x & 61 & 0.20 & 9 \\
         & & ANI1x & 151 & 0.50 & 13 \\
         & & ANI1x + Dispersion & 18 & 0.06 & 5 \\
         & & \textbf{ANIOE62} & \textbf{11} & \textbf{0.04} & \textbf{2}  \\
         \hline
         \multirow{4}{*}{vii} & \multirow{4}{*}{3000}
         & ANI2x & 223 & 0.74 & 25 \\
         & & ANI1x & 68 & 0.23 & 11 \\
         & & ANI1x + Dispersion & 597 & 1.99 & 55 \\
         & & \textbf{ANIOE62} & \textbf{28} & \textbf{0.09} & \textbf{1}  \\
         \hline
         \multirow{4}{*}{viii} & \multirow{4}{*}{30000}
         & ANI2x & 9717 & 3.24 & 976 \\
         & & ANI1x & 16132 & 5.38 & 2545 \\
         & & ANI1x + Dispersion & 12018 & 4.01 & 1242 \\
         & & \textbf{ANIOE62} & \textbf{1727} & \textbf{0.58} & \textbf{193}  \\
         \hline
         \multirow{4}{*}{xii} & \multirow{4}{*}{30000}
         & \textbf{ANI2x} & \textbf{9523} & \textbf{3.17} & \textbf{999} \\
         & & ANI1x & 24267 & 8.09 & 2579 \\
         & & ANI1x + Dispersion & 31852 & 10.62 & 3384 \\
         & & ANIOE62 & 11402 & 3.80 & 1262  \\
         \hline
         \multirow{4}{*}{xvi} & \multirow{4}{*}{30000}
         & ANI2x & 5896 & 1.97 & 644 \\
         & & ANI1x & 6270 & 2.09 & 656 \\
         & & ANI1x + Dispersion & 4787 & 1.60 & 489 \\
         & & \textbf{ANIOE62} & \textbf{549} & \textbf{0.18} & \textbf{71}
         \\

    \end{tabular}
    \caption{Ranking ability of various ANI models for targets with atoms in $\{\text{H, C, N, O}\}$, where a match was found when using random search. Each ANI model was used to rank each population generated by random search, and the Total Rank was computed. The best-performing model for each target is in bold.}
    \label{different_ani_models}
\end{table*}

\begin{table*}[]
    \centering
    \footnotesize
    \begin{tabular}{c|c|c|c|c|c|c}
         Structure & $N_{c}$ & $N_{s}$ & Config & Successful Runs & Min. RMSD [\AA] & Cost [CPU hr] \\\hline
         i & 1 & $10$ & GAtor & 6 & 0.39 & 23.08 \\
         \textbf{i} & \textbf{1} & \textbf{10} & \textbf{symmetric} & \textbf{10} & \textbf{0.26} & \textbf{32.29} \\\hline
         \textbf{ii} & \textbf{1} & \textbf{10} & \textbf{GAtor} & \textbf{4}  & \textbf{0.63} & \textbf{30.30}  \\
         ii & 1 & $10$ & symmetric & 3 & 0.73 & 37.04 \\\hline
         iv & 1 & $10$ & GAtor & 6 & 0.66 & 37.99 \\
         \textbf{iv} & \textbf{1} & \textbf{10} & \textbf{symmetric} & \textbf{7} & \textbf{0.48} & \textbf{61.93} \\\hline
         vi & 5 & $50$ & GAtor & 0 & 0.94 & 96.25 \\
         \textbf{vi} & \textbf{5} & \textbf{50} & \textbf{symmetric} & \textbf{2} & \textbf{0.43} & \textbf{89.92} \\\hline  
         vii & 1 & $10$ & GAtor & 8 & 0.47 & 16.17  \\
         \textbf{vii} & \textbf{1} & \textbf{10} & \textbf{symmetric} & \textbf{10} & \textbf{0.40} & \textbf{23.96} \\\Xhline{3\arrayrulewidth}
         viii & 1 & $100$ & GAtor & 0 & 1.05 & 102.44 \\
         \textbf{viii} & \textbf{1} & \textbf{100} & \textbf{symmetric} & \textbf{4} & \textbf{0.56} & \textbf{93.50} \\\hline  
         x & 1 & $1000000$ & GAtor & 2  & 0.66 & 760.30  \\
         \textbf{x} & \textbf{1} & \textbf{1000000} & \textbf{symmetric} & \textbf{4} & \textbf{0.63} & \textbf{692.80}  \\\hline
         \textbf{xi} & \textbf{1} & \textbf{1000000} & \textbf{GAtor} & \textbf{0} & \textbf{1.10} & \textbf{143.40}  \\
         xi & 1 & $1000000$ & symmetric & 0 & 1.30 & 172.02 \\\Xhline{3\arrayrulewidth}
         xii & 1 & $100$ & GAtor & 3 & 0.74 & 38.95 \\
         \textbf{xii} & \textbf{1} & \textbf{100} & \textbf{symmetric} & \textbf{4} & \textbf{0.59} & \textbf{32.05} \\\hline
         \textbf{xiv} & \textbf{2} & \textbf{200} & \textbf{GAtor} & \textbf{3} & \textbf{0.44} & \textbf{84.36} \\
         xiv & 2 & $200$ & symmetric & 6 & 0.47 & 97.32 \\\Xhline{3\arrayrulewidth}
         xvi & 1 & $100$ & GAtor & 0 & 0.90 & 76.72 \\
         \textbf{xvi} & \textbf{1} & \textbf{100} & \textbf{symmetric} & \textbf{6} & \textbf{0.62} & \textbf{45.97} \\\hline
         xvii & 2 & $200$ & GAtor & 1 & 0.75 & 46.89 \\
         \textbf{xvii} & \textbf{2} & \textbf{200} & \textbf{symmetric} & \textbf{4} & \textbf{0.49} & \textbf{46.92} \\\hline
         xviii & 2 & $200000$ & GAtor & 0 & 1.18 & 393.12  \\
         \textbf{xviii} & \textbf{2} & \textbf{200000} & \textbf{symmetric} & \textbf{0} & \textbf{1.09} & \textbf{281.04} \\\hline
         xix & 60 & $60000$ & GAtor & 0 & 0.92 & 277.02 \\
         \textbf{xix} & \textbf{60} & \textbf{60000} & \textbf{symmetric} & \textbf{3} & \textbf{0.59} & \textbf{244.00} \\\hline
         xx & 10 & $100000$ & GAtor & 0 & 1.55 & 715.75 \\
         \textbf{xx} & \textbf{10} & \textbf{100000} & \textbf{symmetric} & \textbf{2} & \textbf{0.53} & \textbf{503.76} \\\hline
         xxi & 10 & $10000$ & GAtor & 0 & 0.94 & 93.30 \\
         \textbf{xxi} & \textbf{10} & \textbf{10000} & \textbf{symmetric} & \textbf{4} & \textbf{0.57} & \textbf{97.00} \\\Xhline{3\arrayrulewidth}
         xxii & 2 & $200000$ & GAtor & 2 & 0.53 & 101.79 \\
         \textbf{xxii} & \textbf{2} & \textbf{200000} & \textbf{symmetric} & \textbf{4} & \textbf{0.44} & \textbf{135.07} \\\hline
         xxiii-A & 15 & $150000$ & GAtor & 2 & 0.75 & 352.87 \\
         \textbf{xxiii-A} & \textbf{15} & \textbf{150000} & \textbf{symmetric} & \textbf{2} & \textbf{0.70} & \textbf{341.81} \\\hline
         xxiv & 120 & $1200000$ & GAtor & 0 & 1.19 & 410.17 \\
         \textbf{xxiv} & \textbf{120} & \textbf{1200000} & \textbf{symmetric} & \textbf{0} & \textbf{1.05} & \textbf{378.51} \\\hline
         \textbf{xxv} & \textbf{150} & \textbf{150000} & \textbf{GAtor} & \textbf{0} & \textbf{1.40} & \textbf{1213.11} \\
         xxv & 150 & $150000$ & symmetric & 0 & 1.66 & 1021.11 \\\hline
         \textbf{xxvi} & \textbf{3} & \textbf{300000} & \textbf{GAtor} & \textbf{0} & \textbf{1.21} & \textbf{389.05} \\
         xxvi & 3 & $300000$ & symmetric & 0 & 1.24 & 313.19 \\\hline
    \end{tabular}
    \caption{Results when RMSD is used as the objective for selection in the GA. $N_s$ is the number of structures in the initial pool, before filtering and clustering. $N_c$ is the number of molecular conformers considered. The minimum RMSD and computation time required for all 10 runs are also reported. The configuration which achieves the lower minimum RMSD is in bold.}
    \label{table:ga_results}
\end{table*}

\begin{table*}[]
    \centering
    \scriptsize
    \begin{tabular}{c|c|c|c|c|c|c|c|c|c}
         Structure & $N_{c}$ & $N_{s}$ & Model & $N_{mpc}$ & Successful Runs & Min. RMSD [\AA] & Total Rank & Best Rank & Cost [CPU hr] \\\hline
         
         \multirow{2}{*}{i} & \multirow{2}{*}{1} & \multirow{2}{*}{10} 
         & ANI2x & 2,4,8 & 1 & 0.50 & 16514 &  1740 & 152.03 \\
         & & & ANIOE62 & 2,4,8 & 1 & 0.65 & 21655 & 2176 & 218.98 \\\hline
         
         \multirow{2}{*}{ii} & \multirow{2}{*}{1} & \multirow{2}{*}{100} 
         & ANI2x & 4 & 0 & 0.92 & - & - &  26.99 \\
         & & &ANIOE62 & 2,4,8 & 1 & 0.79 & 14086 & 1369 & 106.52 \\\hline
         
         \multirow{2}{*}{iv} & \multirow{2}{*}{1} & \multirow{2}{*}{10} 
         & ANI2x & 2,4,8 & 5 & 0.56 & 45 & 9 & 280.93 \\
         & & & ANIOE62 & 2,4,8 & 7 & 0.33 & 198 & 27 & 405.32 \\\hline
         
         \multirow{2}{*}{vi} & \multirow{2}{*}{5} & \multirow{2}{*}{1000} & 
         ANI2x & 2,4,8 & 3 & 0.50 & 278 &  15 & 794.90 \\
         & & & ANI2OE62 & 2,4,8 & 0 & 0.85 & - & - & 245.37 \\\hline
         
         \multirow{2}{*}{vii} & \multirow{2}{*}{1} & \multirow{2}{*}{10} & ANI2x & 2,4,8 & 10 & 0.24 & 1 & 0 & 109.20 \\
         & & & ANIOE62 & 2,4,8 & 10 & 0.25 & 32 & 4 & 114.65 \\\Xhline{3\arrayrulewidth}
         
         \multirow{2}{*}{viii} & \multirow{2}{*}{1} & \multirow{2}{*}{100} &
         ANI2x & 2,4,8 & 1 & 0.74 & 806 & 76 & 107.86 \\
         & & & ANIOE62 & 2,4,8 & 8 & 0.34 & 244 & 23 & 170.25 \\\hline

         \multirow{2}{*}{x} & \multirow{2}{*}{1} & \multirow{2}{*}{1000000} &
         ANI2x & 4 & 0 & 1.82 & - & -  & 1454.60 \\
         & & & ANIOE62 & 4 & 0 & 2.12 & - & - & 1551.88 \\\Xhline{3\arrayrulewidth}
         
         \multirow{2}{*}{xii} & \multirow{2}{*}{1} & \multirow{2}{*}{100} &
         ANI2x & 8 & 0 & 0.83 & - & - & 30.33 \\
         & & & ANIOE62 & 2,4,8 & 1 & 0.34 & 42 & 0 & 78.47 \\\hline
         
         \multirow{2}{*}{xiv} & \multirow{2}{*}{2} & \multirow{2}{*}{200} & 
         ANI2x & 4 & 0 & 1.19 & - & - & 61.97 \\
         & & & ANIOE62 & 2,4,8 & 1 & 0.40 & 0 & 0 & 524.66 \\\Xhline{3\arrayrulewidth}
         
         \multirow{2}{*}{xvi} & \multirow{2}{*}{1} & \multirow{2}{*}{100} & 
         ANI2x & 8 & 0 & 1.19 & - & - & 54.91 \\
         & & & ANIOE62 & 2,4,8 & 2 & 0.73 & 965 & 95 & 97.60 \\\hline
         
         \multirow{2}{*}{xvii} & \multirow{2}{*}{2} & \multirow{2}{*}{200} & 
         ANI2x & 2,4,8 & 1 & 0.52 & 8688 & 889 & 288.71 \\
         & & & ANIOE62 & 2,4,8 & 1 & 0.33 & 6922 & 722 & 323.08 \\\hline
         
         \multirow{2}{*}{xix} & \multirow{2}{*}{60}& \multirow{2}{*}{60000} & ANI2x & 4 & 0 & 1.26 & - & - & 354.03 \\
         & & & ANIOE62 & 4 & 0 & 1.49 & - & - & 359.87 \\\hline

         \multirow{2}{*}{xx} & \multirow{2}{*}{10} & \multirow{2}{*}{100000} & 
         ANI2x & 4 & 0 & 2.18 & - & - & 779.26 \\
         & & & ANIOE62 & 4 & 0 & 1.91 & -
         & - & 934.67 \\\hline
         
         \multirow{2}{*}{xxi} & \multirow{2}{*}{10} & \multirow{2}{*}{100000} & 
         ANI2x & 4 & 0 & 1.07 & - & - & 154.67 \\
         & & & ANIOE62 & 2,4,8 & 1 & 0.80 & 11375 & 879 & 785.60 \\\Xhline{3\arrayrulewidth}
         
         \multirow{2}{*}{xxii} & \multirow{2}{*}{2} & \multirow{2}{*}{200000} & 
         ANI2x & 4 & 0 & 1.19 & - & - & 165.86 \\
         & & & ANIOE62 & 4 & 0 & 1.63 & - & - & 169.19 \\\hline
         
         \multirow{2}{*}{xxiii-A} & \multirow{2}{*}{15} & \multirow{2}{*}{1500000} &
         ANI2x & 2, 4 & 0 & 1.43 & - & - & 1014.33 \\
         & & & ANIOE62 & 2,4,8 & 1 & 0.42 & 165 &  18 & 2640.80 \\\hline

         \multirow{2}{*}{xxiii-B} & \multirow{2}{*}{15} & \multirow{2}{*}{1500000} &
         ANI2x & 2, 4 & 0 & 1.16 & - & - & 1014.33 \\
         & & & ANIOE62 & 2,4,8 & 1 & 0.44 & 11269 & 1210 & 2640.80 \\\hline

         \multirow{2}{*}{xxiii-C} & \multirow{2}{*}{15} & \multirow{2}{*}{1500000} &
         ANI2x & 2, 4 & 0 & 2.06 & - & - & 1014.33 \\
         & & & ANIOE62 & 2,4,8 & 0 & 1.66 & - &  - & 2640.80 \\\hline

         \multirow{2}{*}{xxiii-D} & \multirow{2}{*}{15} & \multirow{2}{*}{1500000} &
         ANI2x & 2, 4 & 0 & 2.22 & - & - & 1014.33 \\
         & & & ANIOE62 & 2,4,8 & 0 & 2.29 & - &  - & 2640.80 \\\hline

         \multirow{2}{*}{xxiii-E} & \multirow{2}{*}{15} & \multirow{2}{*}{1500000} &
         ANI2x & 2, 4 & 0 & - & - & - & 1014.33 \\
         & & & ANIOE62 & 2,4,8 & 0 & - & - &      - & 2640.80 \\\hline

    \end{tabular}
    \caption{Performance of GAmuza when used with both ANI2x and ANIOE62. For each structure and each $N_{mpc}$ (number of molecules per cell), the entire pipeline is run 10 times to completion. During each run, $N_{s}$ structures are generated for each molecular conformer before filtering and clustering. Successful Runs, Total Rank, Best Rank, minimum RMSD, and computational time are recorded for each target.}
    \label{table:full_results}
\end{table*}

\begin{figure*}[bht]
    \centering
    \includegraphics[width=0.8\linewidth]{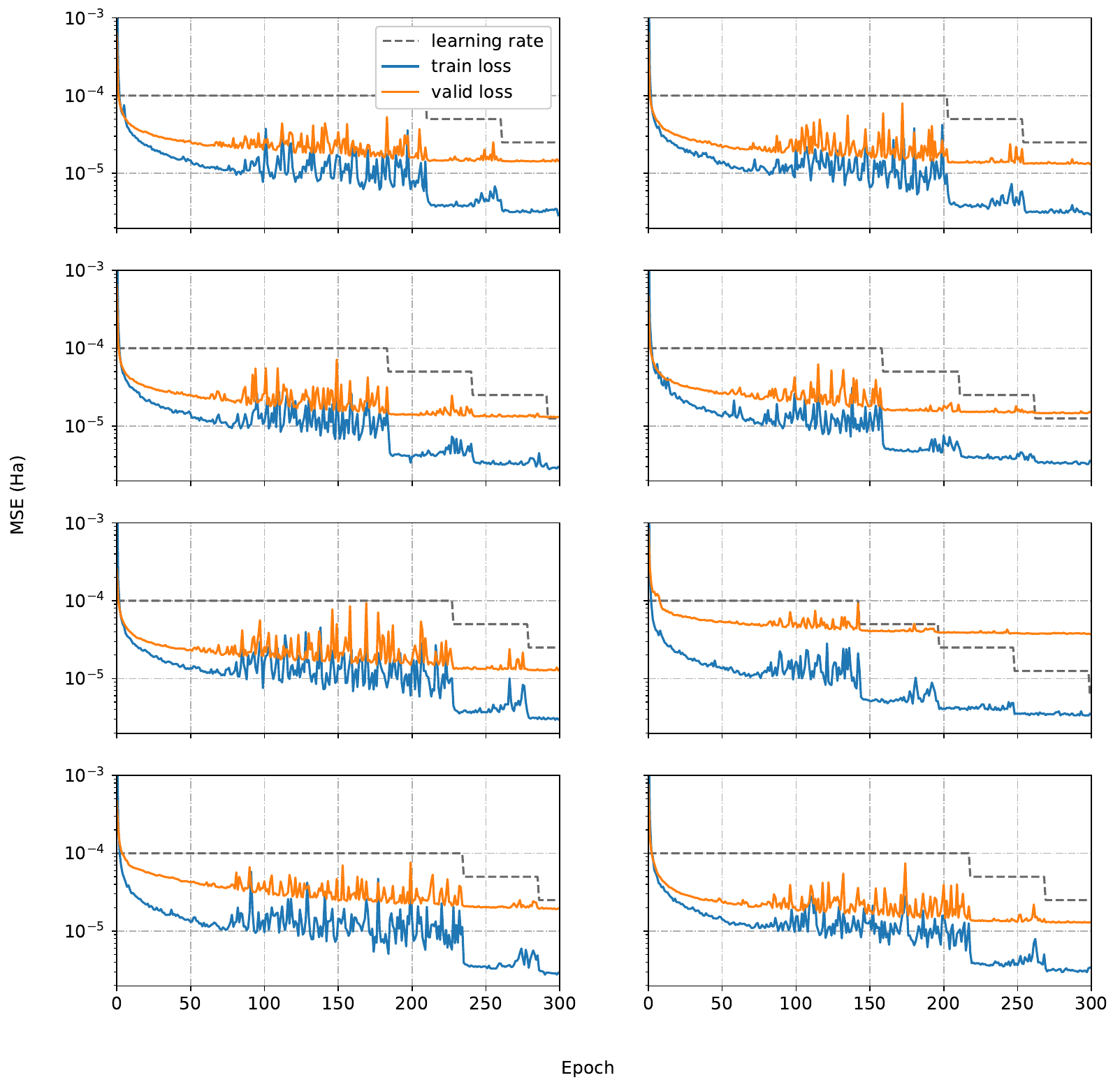}
    \caption{Training and validation curves for training ANIOE62. Losses are calculated using the mean squared error (MSE) of the predicted energy vs. the energy measured at the PBE0 level. Alongside the training and validation losses, the adaptive learning rate is pictured as a dashed line in each plot.}
    \label{fig:matches}
\end{figure*}

\end{document}